\documentclass[epj,nopacs,final]{svjour}
\usepackage{graphics}
\usepackage{latexsym}
\usepackage{subfigure,wrapfig,multirow}
\usepackage{epsfig,color,rotating,amsmath,delarray,array}
\usepackage{makeidx,pifont,float,amssymb}
\definecolor{gray01}{gray}{0.9}
\definecolor{gray02}{gray}{0.8}
\definecolor{gray03}{gray}{0.7}
\definecolor{gray04}{gray}{0.6}
\definecolor{gray05}{gray}{0.5}
\definecolor{gray06}{gray}{0.4}
\definecolor{gray07}{gray}{0.3}
\definecolor{gray08}{gray}{0.2}
\definecolor{gray09}{gray}{0.1}

\begin{document}

\title{\boldmath$\Delta$\unboldmath\ resonances: quark models, chiral symmetry and AdS/QCD}
\titlerunning{$\Delta$ resonances, quark models, chiral symmetry and AdS/QCD}
\authorrunning{E. Klempt {\it et al.}}
\author{Eberhard~Klempt}

\institute{Helmholtz-Institut f\"ur Strahlen- und Kernphysik,
Universit\"at Bonn, Germany}

\abstract{The mass spectrum of $\Delta$ resonances is compared to
predictions based on three quark-model variants, to predictions
assuming that chiral symmetry is restored in high-mass baryon
resonances, and to predictions derived from AdS/QCD. The latter
approach yields a nearly perfect agreement when the confinement
property of QCD is modeled by a soft wall in AdS.
 \vspace*{1mm} \\ {\it PACS:
11.25.Tq Gauge/string duality, 11.30.Rd Chiral symmetries, 12.40.Yx
Hadron mass models and calculations,  14.20.Gk Baryon resonances
with $S=0$} }
\date{Received: \today / Revised version:}

\mail{klempt@hiskp.uni-bonn.de}

\maketitle

\section{Introduction}
The study of high-mass baryon resonances is a demanding task. From
an experimental point of view, most existing information on baryon
resonances \cite{Yao:2006px} has been derived from experiments on
pion (or Kaon) elastic scattering off polarized proton targets which
were performed in the 70ties. Significant data at low pion energies
were added by the pion factories LAMPF, TRIUMF, and PSI (former
SIN). These data were not sufficient to construct the complex
scattering amplitude; experiments at Gatchina have provided a few
measurements of proton recoil polarization and spin rotation
parameter. The data base \cite{Data} is, however, not complete;
analysis and interpretation require further theoretical input from
dispersion relations, analyticity and unitarity. Different analyses
of these data lead to different conclusions concerning mass, width
and the number of observed states. Even worse, the most recent
analysis of Arndt {\it et al.} \cite{Arndt:2006bf} -- which included
more and better data than all analyses before -- found the smallest
number of states.

From a theoretical side, very different concepts have been proposed
to understand the baryon mass spectrum. Quark models -- discussed
here in three variants \cite{Capstick:1986bm,Loring:2001kx,Riska} --
proved to be very successful in explaining the pattern of states
with low-mass excitation energy. At high masses, above 1.8\,GeV,
there is the well known problem of the {\it missing resonances}:
quark models predict many more states than have been observed
experimentally. Possibly, these states have not yet been found as
their conjectured small $N\pi$ couplings prevented their
identification in $\pi N$ elastic scattering. However, they may not
exist. Diquark effects are often invoked to explain the reduced
number of observed states
\cite{Anselmino:1992vg,Kirchbach:2001de,Santopinto:2004hw}.

Glozman observed that in many cases, high-mass baryon resonances
occur in parity doublets \cite{Glozman:1999tk}. He proposed that
their valence quarks could have  typical momenta larger than the
chiral symmetry breaking scale and could decouple from the quark
condensate. If the constituent quark mass originates from their
coupling to the quark condensate, then their constituent (chiral
symmetry breaking) mass should vanish, and chiral symmetry might be
restored in high-mass resonances \cite{Jaffe:2006aq,Glozman:2007jt}.
In this way, high-mass baryon resonances decouple from $N\pi$, in
agreement with experimental observation.

A third approach is based on AdS/QCD, a new development to overcome
the difficulties encountered when dealing with bound states in
quantum chromodynamics. In AdS/QCD the conjectured gauge/gravity
correspondence \cite{Aharony:1999ti} is used to map string modes in
a five-dimensional Anti-de-Sitter (AdS) space into interpolating
`hadronic' operators of a gauge theory defined on its
four-dimensional boundary \cite{Witten:1998qj,Klebanov:2000me}.
Recently, the baryon mass spectrum has been calculated in different
variants of AdS/QCD. The variants differ in the way, conformal
symmetry is broken: either explicitly by an infrared boundary in AdS
(hard wall) \cite{Brodsky:2006uq}, by introducing a non-conformal
dilaton field (called dilation soft wall here), or by infrared
deformation of the AdS metric (metric soft wall)
\cite{Forkel:2007cm,Forkel:2007tz}.

In the preceding paper \cite{Horn:epja}, a study of the reaction
$\gamma p\to p\pi^0\eta$ was presented which led to observation of
$\Delta(1920)P_{33}$ and $\Delta(1940)D_{33}$. Evidence for these
two states had been communicated in a letter \cite{Horn:prl}. In
this paper, we discuss their role for the full spectrum of $\Delta$
resonances and compare $\Delta$ masses with model predictions.
\section{The \boldmath$\Delta$\unboldmath\ mass spectrum and its interpretation}

\subsection{The \boldmath$\Delta$\unboldmath\ mass spectrum}
The lowest-mass state is of course the well known

\begin{center}
\begin{tabular}{ccc}
$\Delta(1232)P_{33}$&&(1)\\
$^{****}$ &\\
\end{tabular}
\vspace{-1mm}\end{center}

\noindent having isospin $I=3/2$ (a property shared by all $\Delta$
resonances), total angular momentum $J=3/2$ and positive parity.
Decays to the $N\pi$ ground state require orbital angular momentum
$L=1$ between pion and nucleon ($P$-wave); the $\Delta(1232)$
quantum numbers are described by $L_{2I,2J}=P_{33}$. In $\pi N$
scattering, it is the first resonance region (which contains a small
$N(1440)P_{11}$ contribution as well). The second resonance region
contains nucleon resonances only, while the third resonance region
receives important contributions from $\Delta$ resonances:

\begin{center}
\begin{tabular}{ccc}
$\Delta(1620)S_{31}$&$\Delta(1700)D_{33}$&(2)\\
$^{****}$&$^{****}$&\\
$\Delta(1750)P_{31}$&$\Delta(1600)P_{33}$&(3)\\
 $^{*}$\hspace{-2mm}&\hspace{-2mm}$^{****}$ &\\
\end{tabular}
\vspace{-2mm}\end{center}

The Particle Data Group (PDG) lists four positive-parity and three
negative-parity states in a narrow mass gap from  1900 to 1950\,MeV.
$\Delta(1950)F_{37}$ is the second resonance (after
$\Delta(1232)P_{33}$) falling onto the main Regge trajectory
comprising $J^P=3/2^+, 7/2^+, 11/2^+, 15/2^+$.\vspace{-2mm}
\begin{center}
\begin{tabular}{ccccc}
$\Delta(1910)P_{31}$\hspace{-2mm}&\hspace{-2mm}$\Delta(1920)P_{33}$\hspace{-2mm}&\hspace{-2mm}$\Delta(1905)F_{35}$\hspace{-2mm}&\hspace{-2mm}$\Delta(1950)F_{37}$\hspace{-2mm}&\hspace{-2mm}(4)\\
$^{****}$ \hspace{-2mm}&\hspace{-2mm} $^{***}$\hspace{-2mm}&\hspace{-2mm}$^{****}$ \hspace{-2mm}&\hspace{-2mm}$^{****}$\\
$\Delta(1900)S_{31}$\hspace{-2mm}&\hspace{-2mm}$\Delta(1940)D_{33}$\hspace{-2mm}&\hspace{-2mm}$\Delta(1930)D_{35}$\hspace{-2mm}&\hspace{-2mm}\hspace{-2mm}&\hspace{-2mm}(5)\\
$^{**}$ \hspace{-2mm}&\hspace{-2mm} $^{*}$\hspace{-2mm}&\hspace{-2mm}$^{***}$\\
\end{tabular}
\vspace{-2mm}\end{center} Intriguingly, $\Delta(1900)$ $S_{31}$,
$\Delta(1940)D_{33}$, and $\Delta(1920)P_{33}$ were not observed in
the most recent partial wave analysis of $\pi N$ data on elastic
scattering by Arndt {\it et al.} \cite{Arndt:2006bf}), while in the
$D_{35}$ wave a state at 2233\,MeV mass and with 773\,MeV width was
found which could be discussed as separate state belonging to the
states (6). Thus different scenarios are possible which will be
discussed at the end of the paper. For the moment we assume that the
seven states (4) and (5) do exist and have masses of about 1.9 to
2.0\,GeV.

The next states were reported with rather weak evidence only. The
PDG lists a $\Delta(2000)F_{35}$ based on two observations giving
masses at about $1735$\,MeV and one with 2200\,MeV; we use the
latter value and call it $\Delta(2190)F_{35}$ to avoid mix-up with
$\Delta(2200)G_{37}$. Surprisingly, $\Delta(2000)F_{35}$ is listed
with two stars. In addition, there are two $\Delta$ candidates with
nega\-tive-parity. None of these states was observed by Arndt {\it
et al.} \cite{Arndt:2006bf}.

\vspace{-2mm}\begin{center}
\begin{tabular}{ccccc}
$\Delta(2150)S_{11}$&\quad-\qquad&-&$\Delta(2200)G_{37}$&\hspace{3.5mm}(6)\\
$^{*}$&&&$^{*}$ &\\
-&\hspace{3.5mm}-&\hspace{2mm}$\Delta(2190)F_{35}$&-&\hspace{3.5mm}(7)\\
 && $^{**}$&&\\
\end{tabular}
\vspace{-2mm}\end{center}

Given the uncertainty of the $\Delta(2190)F_{35}$ mass
determination, this resonance (if it exists at all) could as well be
the missing state of a quartet of positive parity resonances below.
Likewise, two negative-parity states (with quantum numbers $D_{33}$
and $D_{35}$) are missing to form a complete quartet.

The $\Delta(2420)H_{3\,11}$ is an obvious candidate to be a member
of the leading Regge trajectory (with $J=11/2; L=4, S=3/2$).
\vspace{-2mm}
\begin{center}
\begin{tabular}{ccccc}
\qquad-\qquad&$\Delta(2390)F_{37}$&$\Delta(2300)H_{39}$&$\Delta(2420)H_{3\,11}$&(8)\\
& $^{*}$&$^{**}$ &$^{****}$\\
\qquad-\qquad&$\Delta(2360)D_{35}$&-&$\Delta(2400)G_{39}$&(9)\\
 & $^{*}$&&$^{**}$\\
\end{tabular}
\end{center}
It is degenerate in mass with the negative-parity state
$\Delta(2400)G_{39}$. A few further states with positive and
negative parity are known at about the same mass.

Finally we list the two highest mass states, one with negative, one
with positive parity. Likely, they are both members of a quartet of
states. $\Delta(2950)K_{3\,15}$ is the highest-mass state (listed in
RPP) on the leading Regge trajectory.  \vspace{-5mm}
\begin{center}
\begin{tabular}{ccccc}
\qquad\ \ - \  \qquad&\qquad\ \ - \ \qquad&\qquad\ \ - \ \qquad&\quad$\Delta(2750)I_{3\,13}$&(10)\\
&&&$^{**}$ &\\
\qquad\ \ - \  \qquad&\qquad\ \ - \ \qquad&\qquad\ \ - \ \qquad&\quad$\Delta(2950)K_{3\,15}$&(11)\\
&&& $^{**}$&\\
\end{tabular}
\vspace{-2mm}\end{center}

Note the PDG star-rating of the overall status. Only three- and
four-star resonances are considered as established.

\subsection{Quark models}

In quark models, baryons are described by the dynamics of a system
composed of three (constituent) quarks. Assuming locally quadratic
potentials between quarks, wave functions can, to first order in
perturbation theory, be expressed by harmonic-oscillator wave
functions in the two relative coordinates. The spatial wave
functions are characterized by the two orbital angular momenta $l_1$
and $l_2$ and their radial excitation quantum numbers $n_1$ and
$n_2$. The vector sum $\vec l_1+\vec l_2$ defines the total quark
angular momentum $L$. Here, we introduce $\rm L$ as scalar sum of
$l_1$ and $l_2$, ${\rm L}=l_1+l_2$, and  ${\rm N}=n_1+n_2$ as radial
excitation quantum number. Resonances are assigned to a band. The
band number $N$ gives a first estimate of the mass, approximately
the relation $M^2\propto (M^2_{\Delta}+ const \cdot N)$  holds. The
band number is not identical to the radial excitation quantum number
N, instead $N=\rm L+2N$. Baryons are classified by the number $N$ of
excitation quanta, the dimensionality $D$ of the SU(3)
representation, the total angular momentum $J$, the quark orbital
angular momentum $L$, the total quark spin $S$, and by the parity
$P$. Of course, the assignment of intrinsic orbital and spin angular
momenta to the 3 quark system is a non-relativistic concept. There
is no real understanding why the non-relativistic quark model works
so well, but it does work. It is hard to avoid the conclusion that
the five low-mass negative-parity nucleon resonances between 1.5 and
1.7\,GeV

\begin{center}
\begin{tabular}{cccc}
$N(1650)S_{11}$&$N(1700)D_{13}$&$N(1675)D_{15}$\qquad&(12)\\
$^{****}$ &$^{****}$&$^{****}$&\\
$N(1535)S_{11}$& $N(1520)D_{13}$& \ - \ &(13)\\
$^{****}$ &$^{****}$&&
\end{tabular}
\vspace{-2mm}\end{center}

\noindent represent a triplet of states with $(L=1, S=3/2)$ and a
doublet with $(L=1, S=1/2)$. Of course, mixing of states having
identical external quantum numbers but having different spin-orbital
angular momentum configuration is possible. However, the effect of
mixing on the observed masses is small even though mixing may have a
considerable effect on decays where amplitudes and not only
probabilities are relevant. These conclusions are confirmed in
explicit quark model calculations where mixing angles are mostly
small ($<30^{\circ}$).

Quark model predictions and data are compared in Table \ref{model}.
The states are assigned to bands. The ground state
$\Delta(1232)P_{33}$ (1) has $(D,L^P_N)\,S = (56,0^+_0)\,3/2$; it
belongs to a 56-plet in SU(6), total quark spin is $S=3/2$, total
quark angular momentum is $L=0$, and parity is $P=+1$. It belongs to
the ground state baryons, $N=0$. A possible (small) $L=2$ admixture
has attracted high interest \cite{Drechsel:2007sq} but does not
interfere with our conclusions.

The two negative-parity states (2) belong to $(D,L^P_N)\,S =
(70,1^-_1)\,1/2$. The number of observed and expected states
coincides. Two further states (3) similar in mass have positive
parity. The $\Delta(1600)P_{33}$ plays the same role here as the
Roper resonance in the nucleon excitation spectrum. The orbital
angular momenta $l_1=l_2$ vanish and one of the oscillators is
excited radially with $N=1$. It is assigned to $(D,L^P_N)\,S =
(56,0^+_2)\,3/2$. In $\Delta(1750)P_{31}$, both $l_1$ and $l_2$ are
1 and couple to $L=0$ (but $\rm L=2$). For this state, $(D,L^P_N)\,S
= (70,0^+_2)\,1/2$. Both these states are approximately mass
degenerate with the spin doublet $\Delta(1620)S_{31}$ and
$\Delta(1700)D_{33}$ which are members of the first excitation band.
The four states form two parity doublets.

The quartet of states (4) is readily understood as having a total
quark spin $S=3/2$ and total quark angular momentum $L=2$ coupling
to $J=1/2,\cdots,7/2$. The small mass splitting is then interpreted
by nearly vanishing spin-orbit forces. The states are members of the
$(D,L^P_N)\,S = (56,2^+_2)\,3/2$ supermultiplet. The number of
expected states in the second band is 8 while 6 are observed. The
$(D,L^P_N)\,S = (70,2^+_2)\,1/2$ supermultiplet is missing.

\begin{table}[pt]
\caption{\label{model}$\Delta$ resonances, PDG values compared to
quark model predictions. PDG: mass ranges quoted from
\cite{Yao:2006px}. Quark models: negative-parity states in the
$1^{\rm st}$, positive-parity states in the $2^{\rm nd}$,  and
negative-parity states in the $3^{\rm rd}$ excitation band (b.). The
$4^{\rm th}$ and $5^{\rm th}$ band have poorly known states only,
and we do not give model predictions. In \cite{Capstick:1986bm},
states above 2.2\,GeV are omitted; of course, both models predict
the same number of states. A and B refer to two model variants
\cite{Loring:2001kx}. All masses are given in MeV, quark model
predictions are listed in italic.} \begin{center}
\renewcommand{\arraystretch}{1.5}
\hspace{-0.5mm}\begin{tabular}{ccccc} \hline
\hline $1^{\rm st}$ b.&$J^P=1/2^-$&$3/2^-$&\\
\hline
PDG&1630$\pm$30&$1710\pm40$&\\
\hline\cite{Capstick:1986bm}&{\it 1555}\hspace{-1mm}&\hspace{-1mm}{\it 1620}\hspace{-1mm}\vspace{-1mm}\\
\cite{Loring:2001kx},A&{\it 1654}\hspace{-1mm}&\hspace{-1mm}{\it 1628}\hspace{-1mm}\vspace{-1mm}\\
\cite{Loring:2001kx},B&{\it 1625}\hspace{-1mm}&{\it 1633}\hspace{-1mm}\\
\hline\hline $2^{\rm nd}$ b.&$J^P=1/2^+$&$3/2^+$&$5/2^+$&\hspace{-1mm}$7/2^+$\\
\hline
PDG&$\approx1750$&1625$\pm$75&\vspace{-1mm}\\
&1895$\pm$25&1935$\pm$35&\hspace{-3mm}1905$\pm$10&\hspace{-3mm}1930$\pm$20\\
\hline\cite{Capstick:1986bm}&{\it 1835 1875}\hspace{-1mm}&\hspace{-1mm}{\it 1795 1915 1985}\hspace{-1mm}&\hspace{-1mm}{\it 1910 1990}\hspace{-1mm}&\hspace{-1mm}{\it 1940}\vspace{-1mm} \\
\cite{Loring:2001kx},A&{\it 1866 1905}\hspace{-1mm}&\hspace{-1mm}{\it 1810 1871 1950}\hspace{-1mm}&\hspace{-1mm}{\it 1897 1985}&\hspace{-1mm}{\it 1956}\vspace{-1mm}\\
\cite{Loring:2001kx},B&{\it 1901 1928}\hspace{-1mm}&\hspace{-1mm}{\it 1923 1946 1965}\hspace{-1mm}&\hspace{-1mm}{\it 1916 1948}\hspace{-1mm}&\hspace{-1mm}{\it 1912}\\
\hline\hline
$3^{\rm rd}$ b.&$J^P=1/2^-$&$3/2^-$&$5/2^-$&\hspace{-3mm}$7/2^-$\\
\hline
PDG&1900$\pm$50&$\approx1940$&\hspace{-3mm}1960$\pm$60&\vspace{-1mm}\\
&&&\hspace{-2mm}2233$\pm$53$^a$&\hspace{-3mm}$\approx2200$\\
\hline\cite{Capstick:1986bm}&{\it 2035 2140}\hspace{-1mm}&\hspace{-1mm}{\it 2080 2145}\hspace{-1mm}&\hspace{-1mm}{\it 2155 2165}&\hspace{-1mm}{\it 2090}\vspace{-1mm}\\
\cite{Loring:2001kx},A&{\it 2100 2141}\hspace{-1mm}&\hspace{-1mm}{\it 2089 2156 2170}\hspace{-1mm}&\hspace{-1mm}{\it 2170 2187}\hspace{-1mm}&\hspace{-1mm}{\it 2181}\vspace{-2mm}\\
&{\it 2202}&\hspace{-2mm}{\it 2218 2260}&\hspace{-1mm}{\it 2210, 2290}\hspace{-1mm}&\hspace{-1mm}{\it 2239}\vspace{-1mm}\\
\cite{Loring:2001kx},B&{\it 2169 2182}\hspace{-1mm}&\hspace{-1mm}{\it 2161 2177}\hspace{-1mm}&\hspace{-1mm}{\it 2152 2179}\hspace{-1mm}&\hspace{-1mm}{\it 2182}\vspace{-2mm}\\
&{\it 2252}&\hspace{-2mm}{\it 2239 2253 2270}&\hspace{-1mm}{\it 2230 2247}\hspace{-1mm}&\hspace{-1mm}{\it 2220}\\
\hline\hline $4^{\rm th}$ b.&$J^P=5/2^+$&$7/2^+$&$9/2^+$&\hspace{-3mm}$11/2^+$\vspace{-1mm}\\
PDG&2200$\pm$125$^b$&\vspace{-2mm}\\
&&\hspace{-3mm}$\approx$2390&\hspace{-3mm}$\approx$2300&\hspace{-5mm}2400$\pm$100\\
\hline\hline $5^{\rm th}$ b.&$J^P=3/2^-$&$5/2^-$&$7/2^-$&\hspace{-3mm}$9/2^-$\vspace{-1mm}\\
PDG&&\hspace{-3mm}$\approx$2350&&\hspace{-3mm}$\approx$2400\\
\hline\hline
\end{tabular}\end{center}$^a$ from \cite{Arndt:2006bf}, $^b$ from \cite{Cutkosky:1980rh}.
\renewcommand{\arraystretch}{1.0}\vspace{-5mm}
\end{table}

\begin{table*}[pt]\renewcommand{\arraystretch}{1.5}
\caption{\label{chiral}Parity doublets and chiral multiplets of $N$
and $\Delta$ resonances of high mass. List and star rating are taken
from \cite{Yao:2006px}. States not found in the recent analysis of
the GWU group \cite{Arndt:2006bf} are marked by $^a$.}
\begin{center}
\begin{tabular}{ccccccccccc} \hline\hline
$J$=$\frac{1}{2}$&N$_{1/2^+}(2100)^a$&\hspace{-3mm}*&N$_{1/2^-}(2090)^a$&\hspace{-3mm}*&$\Delta_{1/2^+}(1910)$&\hspace{-3mm}****&$\Delta_{1/2^-}(1900)^a$&\hspace{-3mm}**\\
$J$=$\frac{3}{2}$&N$_{3/2^+}(1900)^a$&\hspace{-3mm}**&N$_{3/2^-}(2080)^a$&\hspace{-3mm}**&$\Delta_{3/2^+}(1920)^a$&\hspace{-3mm}***&$\Delta_{3/2^-}(1940)^a$&\hspace{-3mm}*\\
$J$=$\frac{5}{2}$&N$_{5/2^+}(2000)^a$&\hspace{-3mm}**&N$_{5/2^-}(2200)^a$&\hspace{-3mm}**&$\Delta_{5/2^+}(1905)$&\hspace{-3mm}****&$\Delta_{5/2^-}(1930)^a$&\hspace{-3mm}***\\
$J$=$\frac{7}{2}$&N$_{7/2^+}(1990)^a$&\hspace{-3mm}**&N$_{7/2^-}(2190)$&\hspace{-3mm}****&$\Delta_{7/2^+}(1950)$&\hspace{-3mm}****&$\Delta_{7/2^-}(2200)^a$&\hspace{-3mm}*\\
$J$=$\frac{9}{2}$&N$_{9/2^+}(2220)$&\hspace{-3mm}****&N$_{9/2^-}(2250)$&\hspace{-3mm}****&$\Delta_{9/2^+}(2300)$&\hspace{-3mm}**&$\Delta_{9/2^-}(2400)^a$&\hspace{-3mm}**\\
\hline\hline
\end{tabular}
\vspace*{1mm}
\renewcommand{\arraystretch}{1.0}
\end{center}
\end{table*}

Likewise it is tempting to interpret the negative parity states (5)
as $L=1, S=3/2$ states, coupling to $J=1/2,\,3/2,\,5/2$, again with
vanishing spin-orbit forces. $S=3/2$ $\Delta$ states are symmetric
in their spin and flavor wave function, antisymmetric in their color
wave function, hence their spatial wave function must be symmetric.
This is impossible for the $L=1$ ground state. For a spatially
symmetric state, at least one of the two oscillators of the
three-body system must be radially excited. If the states (5) are
interpreted as spin triplet of resonances, they must have $L=1, N=1,
S=3/2$ and belong to the $(D,L^P_N)\,S = (56,1^-_3)\,3/2$
supermultiplet. The positive and negative parity states are mass
degenerate, they form three parity doublets; only
$\Delta(1950)F_{37}$ is not accompanied by an odd-parity partner. In
the $3^{\rm rd}$ band, 14 states are expected while at most 5 are
observed. The missing supermultiplets have $(D,L^P_N)\,S =
(70,1^-_3)\,1/2$, $(70,1^-_3)\,1/2$, $(70,2^-_3)\,1/2$, and
$(56,3^-_3)\,3/2$.

$\Delta(2200)G_{37}$ (6) is likely member of a $(D,L^P_N)\,S$ =
$(70,3^-_3)$ $1/2$ spin doublet with its $J^P=5/2^-$ companion
missing. Its partner $\Delta(2150)S_{31}$ is the third resonance in
this partial wave. While $\Delta(1900)S_{31}$ is rather low in mass,
$\Delta(2150)S_{31}$ fit very well to quark models. The two
states~(6) are mass degenerate with $\Delta(2190)F_{35}$ (7). Quark
models predict two $\Delta F_{35}$ states  in the second excitation
band but the $\Delta(1905)F_{35}$--$\Delta(2190)F_{35}$ mass gap is
much larger than expected. Hence we rather prefer to assign
$\Delta(2190)F_{35}$ (at 2200\,MeV) to the forth excitation band
with $L=2$, N=1.

The three positive-parity and two negative-parity states in (8) and
(9) also fall into a narrow mass window. The two states
$\Delta(2400)G_{39}$ and $\Delta(2420)H_{3\,11}$ are best
interpreted assuming $S=3/2$ and $L=3$ or $L=4$, respectively. If
they had total quark spin $S=1/2$, they would need to have two more
units of orbital angular momentum which would place them higher in
mass. Of course, a small admixture of high angular momenta is
possible but does not affect this discussion. We assign the states
in (8) to the forth excitation band while those in (9) might belong
to the fifth band. The number of expected states increases
considerably; we refrain from assigning the few observed states to
supermultiplets. The two final states (10,11) need both spin $S=3/2$
to form, with $L=5$ and $6$, respectively, the observed total
angular momentum.

The predictions of quark models (see Table \ref{model}) agree only
partly with observations. First, the number of expected states is
much larger than observed. This is the well known problem of missing
resonances. The second discrepancy concerns the mass pattern. Quark
models predict a clear separation of states belonging to different
excitation bands. Instead, states belonging to different bands are
often {\it degenerate in mass}. The states of negative parity in the
first excitation band are mass degenerate with states listed in the
first line of the second band (Table \ref{model}). The second line
in the $2^{\rm nd}$ band is mass degenerate with the first line in
the $3^{\rm rd}$ band, and this pattern continues even though with
decreasing reliability of data and interpretation.

Disagreement is found whenever we have assigned a unit of radial
excitation to a resonance. The negative parity states in the first
excitation band (2) are mass degenerate with the positive parity
states in (3) belonging to the second band. $\Delta(1600)P_{33}$ is
a radial excitation, $\Delta(1750)P_{31}$ has intrinsically orbital
excitations and should have higher mass, too.  We have argued that
the triplet of negative-parity states at about 1920\,MeV must have
$N=1$ and belong to the $3^{\rm rd}$ band; they are mass degenerate
with states belonging to the second band. $\Delta(2400)$ $G_{39}$
from the $5^{\rm th}$ band is mass degenerate with
$\Delta(2420)H_{3\,11}$ of the $4^{\rm th}$ band. In all cases, the
problem arises since a radial excitation to $\rm N=1$ corresponds to
a change of the band number $N$ by two units, $N=\rm L+2N$. Yet
experimentally, states of opposite parities acquire the same mass
which suggests $M^2\propto\rm L+N$. Two reasons have been proposed
for the pattern. The first one which we discuss next is restoration
of chiral symmetry in high-mass baryon resonances.

\subsection{Parity doublets from chiral symmetry restoration}
The parity doublets were interpreted by Glozman
\cite{Glozman:1999tk} as evidence for restoration of chiral symmetry
in high-mass excitations. There is an abundant literature on this
subject, see \cite{Jaffe:2006jy,Glozman:2007ek} for two recent
reviews. Depending on how the symmetry is realized in Nature, parity
doublets must not interact by pion emission or absorption, a
striking prediction that can be tested experimentally
\cite{Jaffe:2006aq,Glozman:2007jt}. The weakness of the signals for
high-mass baryons in elastic $\pi N$ scattering may thus provide
further evidence for restoration of chiral symmetry.

\begin{figure*}[pt]
\centerline{
\epsfig{file=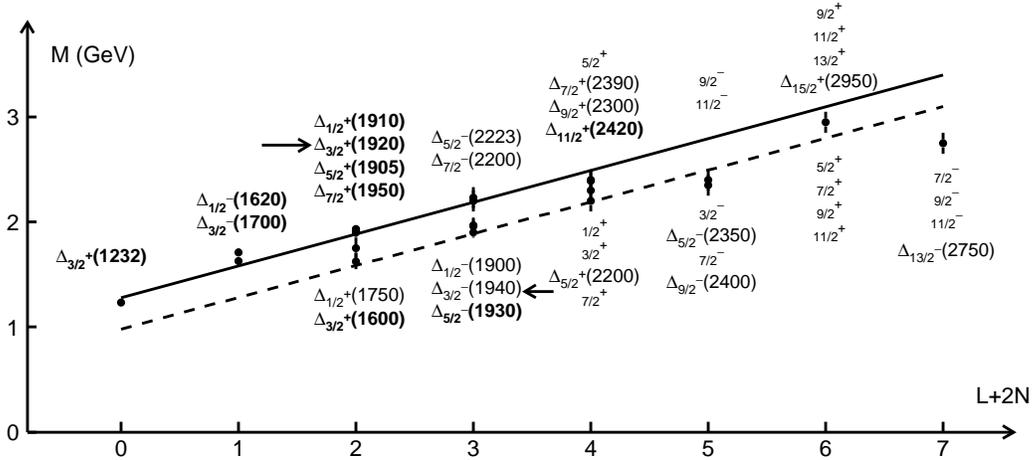,width=0.8\textwidth,clip=}}
\caption{\label{delta_hard} Mass of $\Delta$ resonances as a
function of the leading intrinsic orbital angular momentum $L$ and
the radial excitation quantum number $N$ (corresponding to $n_1+n_2$
in quark models). The line represents a prediction based on the
AdS/QCD hard wall model \cite{Brodsky:2006uq}. Resonances assigned
to $N=0$ and $N=1$ are listed above or below the trajectory. The two
states reported here are indicated by arrows. The solid line should
connect $\Delta$'s belonging to a 56-plet, the dashed line should
connect states belonging to a 70-plet. }
\end{figure*}

Table \ref{chiral} presents high-mass $N$ and $\Delta$ resonances
\cite{Yao:2006px}. The PDG star-rating is also given. Among the 10
parity doublets there are just 2 doublets for which both partners
can be considered as established (with both partners having 3 or 4
stars). Reliable information on all four states of a chiral
multiplet exists in no case. Even worse, a recent analysis of Arndt
{\it et al.} \cite{Arndt:2006bf} implementing recent precise data on
elastic $\pi N$ scattering from meson factories did not find any of
the states with 1 or 2 stars, and just one of the parity doublets
survives.

It is obvious that a new experimental approach beyond elastic $\pi
N$ scattering is needed to explore the high-mass region of $N$ and
$\Delta$ resonances with such weak couplings to the $\pi N$ channel.
Photoproduction of multiparticle final states seems to be a good
choice to avoid $\pi N$ in both the initial and the final state.
Recently, evidence for one full $J=3/2$ chiral multiplet was
reported with all four states derived from photoproduction. The
preceding paper described the reaction $\gamma p\to p\pi^0\eta$ from
which the existence of $\Delta(1920)P_{33}$ and $\Delta(1940)D_{33}$
was deduced. The masses were determined to ($1980^{+25}_{-45}$) and
($1985\pm 30$)\,MeV, respectively. In \cite{Nikonov:2007br},
evidence was presented for $N(1900)P_{13}$, at $(1915\pm50)$\,MeV,
from an analysis of a large variety of photo- and pion-induced
reactions, in particular from the new CLAS measurements of double
polarization observables for photoproduction of hyperons. The forth
resonance, a $N(1875)D_{13}$, was discussed as SAPHIR resonance in
the literature \cite{Bennhold:1998ib}. In \cite{Sarantsev:2005tg},
its mass was determined to $(1875\pm25)$\,MeV. The four states can
be interpreted as two parity doublets, a $N$ doublet at 1900\,MeV
and a $\Delta$ doublet at 1980\,MeV. As full chiral multiplet, mass
breaking effects of the order of 80\,MeV have to be accepted.

From the experimental side, there is one problem with the concept of
chiral symmetry restoration in high-mass resonances which was first
pointed out in \cite{Klempt:2002tt}: the absence of a near-by parity
partner of states like $\Delta(1950)F_{37}$. Quite in general,
``stretched" states with the maximum angular momentum $J=L+S$ and
$L$ even ($\Delta(1232)P_{33}$, $\Delta(1950)F_{37}$,
$\Delta(2420)H_{3\,11}$, $\Delta(2950)K_{3\,11}$) seem to have no
parity partner. The need to invoke chiral symmetry restoration to
explain parity doublets was thus disputed \cite{Klempt:2007cp}. In
view of the weak status of most high-mass resonances,
non-observation of ``stretched" states is of course not a conclusive
argument.

\subsection{Baryon resonances from AdS/QCD}

\begin{figure*}[pt]
\centerline{
\epsfig{file=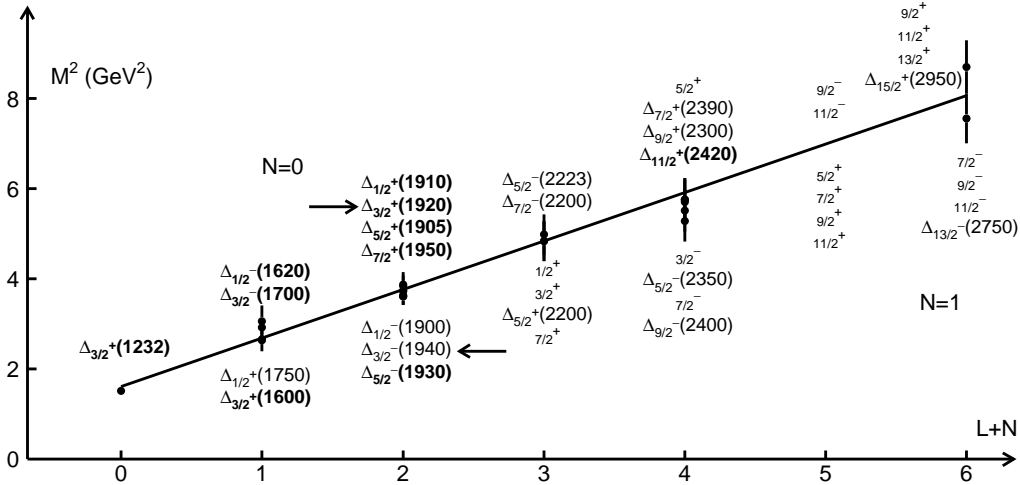,width=0.8\textwidth,clip=}}
\caption{\label{delta_table} Regge trajectory for $\Delta$
resonances as a function of the leading intrinsic orbital angular
momentum $L$ and the radial excitation quantum number $N$
(corresponding to $n_1+n_2$ in quark models). The line represents a
prediction  of the metric-soft-wall AdS/QCD model
\cite{Forkel:2007cm,Forkel:2007tz}. Resonances with $N=0$ and $N=1$
are listed above or below the trajectory. The mass predictions are
1.27, 1.64, 1.92, 2.20, 2.43, 2.64, 2.84\,GeV. The two states
reported in \cite{Horn:epja,Horn:prl} are indicated by arrows. }
\end{figure*}

The degeneracy of positive and negative parity states in Table
\ref{model} is phenomenologically reproduced if $N={\rm L+N}$ were
approximately correct, a relation which is not supported by quark
models, as can be seen from Table \ref{model}. Radial excitations
with $\rm N>0$ are not found at masses corresponding to the
harmonic-oscillator band $N=\rm L+2N$ but rather to a band defined
by $N={\rm L+N}$. This relation -- the mass of excited baryons is a
function of $\rm L+N$ and not of $\rm L+2N$ -- was recently derived
within a new holographic approach which analytically solves an
approximate version of QCD where the strong coupling is large and
nearly constant. (In fact, the QCD coupling itself may become
scale-independent if its renormalization group flow approaches a
conformal fixed point in the infrared. Dimensional counting rules
for the near-conformal power law fall-off \cite{Lepage:1979za} can
be used to argue that at large distances, the strong coupling
constant indeed approaches a constant \cite{Brodsky:2008pg}.)

The AdS/QCD approach is based on a (conjectured) equivalence between
a string theory defined in a spacetime which becomes Anti-de-Sitter
(AdS) near its boundary, and a gauge theory defined on this
boundary. The gauge/string correspondence then relates interpolating
operators (including those which carry hadronic quantum numbers) of
the boundary gauge theory (and their correlators) to the propagation
of weakly coupled strings in the five- dimensional, asymptotically
AdS space \cite{Witten:1998qj,Klebanov:2000me}. Hence AdS/QCD bares
the promise to simultaneously describe the nucleon mass spectrum and
the partonic degrees of freedom observed in deep inelastic
scattering \cite{Brodsky:2007hb}.

In AdS/QCD the confinement properties of QCD are linked to infrared
modifications of the asymptotically AdS space. Several variants have
been proposed which differ in how they implement confinement. In
`hard wall' models (see \cite{Polchinski:2001tt} and references
therein) a cutoff is imposed on the fifth AdS dimension and
generates spectra of the type $M_N(L) \propto \rm L + 2N$
\cite{Brodsky:2006uq,Brodsky:2008pg}. In the `dilaton soft wall'
model [16] an additional bulk dilaton field breaks conformal
symmetry and generates confinement effects in the meson sector. In
the `metric soft wall' model \cite{Karch:2006pv}, finally, smooth IR
deformations of the AdS metric itself implement confinement and
generate baryon spectra of the form $M^2_N (L) \propto\rm L + N$
\cite{Forkel:2007cm,Forkel:2007tz}. In refs.
\cite{Brodsky:2006uq,Forkel:2007cm} only three- and four-star
resonances were used to compare the predictions with data.

In Fig. \ref{delta_hard}, adapted from \cite{Brodsky:2006uq}, not
only established resonances but also one- and two-star resonances
are included. The main (solid) line represents $\Delta$ resonances
with intrinsic spin 3/2, the dashed line those having $S=1/2$. The
splitting between the two different spin configurations was seen to
be required for nucleons where `good diquarks'  (with spin and
isospin both equal to zero) make the resonance lighter compared to
nucleons with `bad diquarks' (with $S=1$ or $I=1$). In decuplet
states there are, of course, only bad diquarks.

At low masses there are a few states falling onto the wrong
trajectory. The two negative parity states $\Delta(1620)$ $S_{31}$
and $\Delta(1700)D_{33}$ must have $S=1/2$ and should fall onto the
dashed line; yet they have masses just on the solid line. Of the two
positive-parity states, $\Delta(1600)P_{33}$ has $S=3/2$ and is
placed on the dashed line; its should fall onto the solid line.
$\Delta(1750)P_{31}$ has $S=1/2$, should have a lower mass than
$\Delta(1600)P_{33}$ and be found on the dashed line.

In \cite{Brodsky:2006uq}, $\Delta(1930)D_{35}$ was interpreted as
$L=3, S=1/2$ excitation. The new evidence for $\Delta(1940)D_{33}$
-- which seems to be a natural spin partner of $\Delta(1930)D_{35}$
-- suggests $L=1,S=3/2,N=1$ quantum numbers for both, and the
two-star $\Delta(1900)S_{31}$ to be the natural third partner to
complete a spin triplet. In the interpretation of
\cite{Brodsky:2006uq}, one could of course also argue that
$\Delta(1900)S_{31}$ and $\Delta(1940)D_{33}$ have $L=1,S=1/2,\rm
N=1$, and $\Delta(1930)D_{35}$ and a missing $\Delta G_{37}$ below
2\,GeV are $L=3, S=1/2$ excitations.

At high masses, some problems remain. In particular
$\Delta(2750)I_{3\,13}$ is far from the solid line.

In conclusion, there are clear discrepancies between hard-wall
AdS/QCD and data in the 1.7\,GeV region. Above 1.8\,GeV, some
inconsistencies with the hard wall solution exist, in particular the
existence of $\Delta(1940)D_{33}$ \cite{Horn:epja,Horn:prl} and the
non-observation of a $\Delta G_{37}$ candidate with mass between 1.9
and 2\,GeV are difficult to reconcile with hard-wall AdS/QCD. But
overall, the trend of most established states is reasonably
reproduced.

In \cite{Forkel:2007cm,Forkel:2007tz}, the mass spectrum of light
mesons and baryons was predicted using AdS/QCD in the metric
soft-wall approximation. Relations between ground state masses and
trajectory slopes
\begin{eqnarray}
M^2 &=& 4\lambda^2 \rm (L+N+1/2)\qquad {\rm for\ mesons}\nonumber\\
M^2 &=& 4\lambda^2 \rm (L+N+3/2)\qquad {\rm for\ baryons}\quad  (A)
\vspace{-4mm}\nonumber\end{eqnarray} were derived. Using the slope
of the $\Delta$ trajectory, masses were calculated. They are plotted
as a function of $\rm L+N$ in Fig. \ref{delta_table}. The two states
indicated by arrows are those found in \cite{Horn:epja,Horn:prl}.
While the positive-parity $\Delta(1920)P_{33}$ has three stars in
the PDG rating, the negative-parity $\Delta(1940)D_{33}$ had one
star only. Both states were not observed in the latest analysis of
Arndt {\it et al.} \cite{Arndt:2006bf} on elastic $\pi N$
scattering.

The four positive- and negative-parity states between 1.60 and
1.75\,GeV (2,3) are predicted to have the same mass
(1.62\,GeV){\footnote{\footnotesize The $\Delta_{1/2^+}(1750)$ is
tricky; it has $\rm L=2$ but both oscillators are excited. Since
they are orthogonal, the internal separations increase less than for
parallel angular momenta.}}; the seven states (4,5) should have 1.92
GeV. The predicted masses for $\rm L+N=3$ (6,7) and 4 (8,9) are 2.20
and 2.42\,GeV, respectively. The trajectory continues with the
calculated masses 2.64 for $\rm L+N=5$ and 2.84\,GeV for $\rm
L+N=6$. Experimentally, the highest mass state is
$\Delta(2950)K_{3\,15}$ which requires $\rm L=6$. In this
interpretation, $\Delta(2750)I_{3\,13}$ has $\rm L=5, S=3/2$ and
$\rm N=1$ and should be degenerate in mass with
$\Delta(2950)K_{3\,15}$. Both are expected to have a mass of
2.84\,GeV which is not incompatible with the experimental findings
even though the mass difference of 200\,MeV between the two states
does not support their expected mass degeneracy.

An early interpretation of strings was proposed by Nambu
\cite{Nambu:1978bd}. He assumed that the gluon flux between the two
quarks is concentrated in a rotating flux tube or a rotating string
with a homogeneous mass density. Nambu derived a linear relation
between squared mass and orbital angular momentum, $M^2\propto L$.
This mechanical picture was further developed  by Baker and Steinke
\cite{Baker:2002km} and by Baker \cite{Baker:2003fn} to a field
theoretical approach. For mesons, the functional dependence ($A$)
was derived.

The relation (A) between $\Delta$ masses and $L$ and $N$ has been
derived earlier in a phenomenological analysis of the baryon mass
spectrum \cite{Klempt:2002vp}.  For octet and singlet baryons, one
term ascribed to instanton-induced interactions was required to
reproduce the full mass spectrum of all baryon resonances having
known spin and parity.

The striking agreement between the  measured baryon excitation
spectrum and the predictions \cite{Forkel:2007cm,Forkel:2007tz}
based on AdS/QCD and the success of the phenomenological mass
formula \cite{Klempt:2002vp} pose new questions. In both cases, the
baryon masses depend on the number of orbital and radial excitations
just as mesons. But baryons have an extra degree of freedom. So,
where are the hidden states and, why are these states not realized
in Nature? Why do they not appear in predictions of AdS/QCD\,? The
active programs at several laboratories to pursuit baryon
spectroscopy using photon beams of linear and circular polarization
and polarized target underline the hope that these issues can be
solved in future.

\subsection{Alternative interpretations}
The experimental situation is, unfortunately, unsettled. The
$\Delta$ spectrum in the 1900\,MeV region plays a decisive role in
interpretation. At the end, we discuss briefly a few alternative
scenarios which could be true. We discuss the possibility that 1.,
 $\Delta(1930)D_{35}$ does not exist, or 2., $\Delta(1900)S_{31}$
does not exist or 3., that $\Delta G_{37}$ exists in this mass range
but has not yet been found.

\begin{itemize}
\item[1.] The two resonances $\Delta(1900)S_{31}$ and $\Delta(1940)D_{33}$
exist at about their nominal masses but the $\Delta(1930)D_{35}$ has
a mass of about 2233\,MeV as found in \cite{Arndt:2006bf}.
\end{itemize}

In this case, $\Delta(1900)S_{31}$ and $\Delta(1940)D_{33}$ would
form a super-doublet of (dominantly) $L=1,S=1/2$ resonances and are
likely radial excitations with $L=1,S=1/2$ resonances and could
belong to one of the two $J^P\,(D^P_N)\,S = (70,1^-_3)\,1/2$
supermultiplets. The masses do not agree well with quark model
calculations \cite{Capstick:1986bm,Loring:2001kx} (see Table
\ref{model}) which predict a $\approx 180$\,MeV mass gap between the
negative-parity  and the positive-parity states. However, given the
experimental uncertainties, the experimental masses are not
completely incompatible with the quark models. The
$\Delta(2223)D_{35}$ falls into the mass range where it is predicted
in all three quark models. With its large width, it may comprise
several resonances. This scenario is in mild disagreement with quark
models, disagrees with hard-wall and supports the metric soft-wall
version of AdS/QCD.

\begin{itemize}
\item[2.] The $\Delta(1900)S_{31}$ may not exist at about its nominal
mass but both, $\Delta(1940)D_{33}$ and $\Delta(1930)D_{35}$ exist.
\end{itemize}

In this case, $L=1$ is unlikely. The two states could indicate a
spin doublet with $L=2$; the two intrinsic oscillators could have
orbital angular momenta $l_1=2$ and $l_1=1$ coupling to $L=2$. Such
configurations have never been observed so far and thus, this
possibility would be very exciting. It would be a striking
confirmation of quark models.

Alternatively, the $\Delta(1940)D_{33}$ could be partner of a spin
quartet with $L=3$ which would include $\Delta(2400)G_{39}$ and two
unobserved states with $J=7/2$ and $5/2$. The large mass difference
makes this possibility rather unlikely.

\begin{itemize}
\item[3.] The three resonances $\Delta(1900)S_{31}$, $\Delta(1940)D_{33}$,
$\Delta(1930)$ $D_{35}$ exist at about their nominal masses. In
addition, a $\Delta G_{37}$ exists in this mass range but has not
yet been found.
\end{itemize}
The four resonances could form two doublets of (dominantly)
$L=1,S=1/2,N=1$ and $L=3,S=1/2,N=0$. Their masses are uncomfortably
low when compared to quark models. The four negative-parity
resonances would be accompanied by four positive-parity partners,
$\Delta(1910)P_{31}$, $\Delta(1920)P_{33}$, $\Delta(1905)F_{35}$,
and $\Delta(1950)F_{37}$. This scenario would be a striking
confirmation of the scenario of chiral symmetry restoration in which
parity doublets are predicted for all high-mass resonances.

\section{Summary} The confirmation of two $\Delta$
resonances \cite{Horn:epja,Horn:prl} of doubtful existence has
initiated a comparison of the experimental $\Delta$ mass spectrum
with model predictions. Quark models predict a much larger number of
states than observed in experiments; this is the well known problem
of missing resonances. Quark models also predict radial excitations
to have higher masses than observed experimentally, the Roper (in
the nucleon excitation spectrum) is the best known example. A large
number of states exist which can be grouped pairwise into parity
doublets. The two states confirmed in \cite{Horn:epja,Horn:prl} form
a parity doublet as well. Chiral symmetry restoration predicts the
existence of further states.

The coincidence between masses (and abundance) of known $\Delta$
resonances and very simple mass relations derived in AdS/QCD is
intriguing. In particular when the confinement of QCD is modeled by
a soft infrared deformation of the AdS metric, there is striking
agreement between data and the prediction. The masses of all 23
$\Delta$ resonances are well reproduced by just one single
parameter, the slope of the Regge trajectory.

Does this success imply that AdS/QCD and the string picture are
right and quark models and the concept of chiral symmetry
restoration are wrong\,? We do not believe so. AdS/QCD and the
string picture pick up an important aspect of the baryon spectrum,
the treatment of confinement. In AdS/QCD, confinement is
parameterized as a `soft' (or `hard') limit for the off-shell
structure of quark dynamics. This seems to work better than using a
linear confinement potential used in quark models. But AdS/QCD does
not tell us why we have five low-mass negative-parity resonances of
the nucleon (12,13) and just two of the $\Delta(1232)$ (2). And
there could be a connection between two experimental observations:
the `stretched' states with $J=L+S$ are those which are best seen in
pion elastic scattering experiments and they are those which miss a
parity partner. Hence possibly, chiral doublets develop only for
states weakly coupled ton $\pi N$. Likely, different models pick up
different aspects of the baryon spectrum. Certainly, we are still
far from a complete understanding of the dynamics of the formation
of baryon resonances.

More data are needed to confirm or to disprove the present findings;
to arrive at a solid understanding of the complicated pattern of
highly excited $N$ and $\Delta$ resonances, intense efforts are
mandatory, in experiments, partial wave analyses and in theoretical
foundations.

\section*{Acknowledgement}
This paper was initiated by new results of the CBELSA collaboration;
grateful thanks go to the collaboration for many fruitful
discussions. In particular I would like to thank B. Krusche for a
critical reading of the manuscript. I would like to thank
S.J.~Brodsky and G.F.~de Teramond for illuminating discussions on
AdS/QCD and its relation to baryon spectroscopy. I had the
opportunity to discuss the different AdS/QCD approaches with H.
Forkel; clarifying discussions and suggestions for the correct
wording to describe AdS/QCD are gratefully acknowledged. I am
indebted to B. Metsch and H. Petry for numerous discussions on the
quark model.


\begin{thebibliography}{20}
\bibitem{Yao:2006px}
  W.M.~Yao {\it et al.},
  J.\ Phys.\ G {\bf 33} (2006) 1 and 2007 partial update for the 2008 edition.
\bibitem{Data}The database can be accessed at SAID facility at the website
http://gwdac.phys.gwu.ed
\bibitem{Arndt:2006bf}
  R.~A.~Arndt, W.J.~Briscoe, I.I.~Strakovsky and R.L.~Workman,
  Phys.\ Rev.\  C {\bf 74} (2006) 045205.
\bibitem{Capstick:1986bm}
S.~Capstick, N.~Isgur,
Phys.\ Rev.\ D {\bf 34} (1986) 2809.
\bibitem{Loring:2001kx}
U.~L\"oring, B.~C.~Metsch and H.~R.~Petry,
Eur.\ Phys.\ J.\ A {\bf 10} (2001) 395, 447.
\bibitem{Riska}L.~Y.~Glozman, W.~Plessas, K.~Varga and R.~F.~Wagenbrunn,
Phys.\ Rev.\ D {\bf 58} (1998) 094030.
\bibitem{Anselmino:1992vg}
  M.~Anselmino, E.~Predazzi, S.~Ekelin, S.~Fredriksson and D.B.~Lichtenberg,
  Rev.\ Mod.\ Phys.\  {\bf 65} (1993) 1199.
\bibitem{Kirchbach:2001de}
  M.~Kirchbach, M.~Moshinsky and Yu.~F.~Smirnov,
  Phys.\ Rev.\  D {\bf 64} (2001) 114005.
\bibitem{Santopinto:2004hw}
  E.~Santopinto,
  Phys.\ Rev.\  C {\bf 72} (2005) 022201.
\bibitem{Glozman:1999tk}
  L.Y.~Glozman,
  Phys.\ Lett.\  B {\bf 475} (2000) 329.
\bibitem{Jaffe:2006aq}
  R.L.~Jaffe, D.~Pirjol and A.~Scardicchio,
  Phys.\ Rev.\  D {\bf 74} (2006) 057901.
\bibitem{Glozman:2007jt}
 See, e.g., L.Y.~Glozman,
  Phys.\ Rev.\ Lett.\  {\bf 99} (2007) 191602, and references
  therein.
\bibitem{Aharony:1999ti}
  O.~Aharony, S.~S.~Gubser, J.~M.~Maldacena, H.~Ooguri and Y.~Oz,
  Phys.\ Rept.\  {\bf 323} (2000) 183.
\bibitem{Witten:1998qj}
  E.~Witten,
  Adv.\ Theor.\ Math.\ Phys.\  {\bf 2} (1998) 253.
\bibitem{Klebanov:2000me}
  I.R.~Klebanov,
  ``TASI lectures: Introduction to the AdS/CFT correspondence,''
  arXiv:hep-th/0009139.
  \bibitem{Karch:2006pv}
  A.~Karch, E.~Katz, D.~T.~Son and M.~A.~Stephanov,
  Phys.\ Rev.\  D {\bf 74} (2006) 015005.
\bibitem{Brodsky:2006uq}
  S.J.~Brodsky,
  Eur.\ Phys.\ J.\  A {\bf 31} (2007) 638.
\bibitem{Forkel:2007cm}
  H.~Forkel, M.~Beyer and T.~Frederico,
  JHEP {\bf 0707} (2007) 077.
\bibitem{Forkel:2007tz}
  H.~Forkel, M.~Beyer and T.~Frederico,
  Int.\ J.\ Mod.\ Phys.\  E {\bf 16} (2007) 2794.
\bibitem{Horn:epja}
I. Horn {\it et al.}, preceding paper.
\bibitem{Horn:prl}
I. Horn {\it et al.}, ``Evidence for a parity doublet
$\Delta(1920)P_{33}$ and $\Delta(1930)D_{33}$ from $\gamma p\to
p\pi^0\eta$", arXiv:.
\bibitem{Drechsel:2007sq}
  D.~Drechsel and T.~Walcher,
  ``Hadron structure at low Q$^2$,''
  arXiv:0711.3396 [hep-ph].
   \bibitem{Cutkosky:1980rh}
  R.E.~Cutkosky , C.P.~Forsyth, J.B.~Babcock, R.L.~Kelly and R.E.~Hendrick,
  ``Pion - Nucleon Partial Wave Analysis,''
 $4^{\rm th}$ Int. Conf. on Baryon Resonances, Toronto, Canada, July
 14-16, 1980. QCD161:C45:1980.
\bibitem{Jaffe:2006jy}
  R.L.~Jaffe, D.~Pirjol and A.~Scardicchio,
  Phys.\ Rept.\  {\bf 435} (2006) 157.
\cite{Glozman:2007ek}
\bibitem{Glozman:2007ek}
  L.~Y.~Glozman,
  Phys.\ Rept.\  {\bf 444} (2007) 1.
\bibitem{Nikonov:2007br}
  V.~A.~Nikonov, A.~V.~Anisovich, E.~Klempt, A.~V.~Sarantsev and U.~Thoma,
  Phys.\ Lett.\  B {\bf 662} (2008) 245.
\bibitem{Bennhold:1998ib}
  C.~Bennhold, T.~Mart, A.~Waluyo, H.~Haberzettl, G.~Penner, T.~Feuster and U.~Mosel,
  ``Nucleon resonances in kaon photoproduction,''
 Workshop on Electron Nucleus Scattering, Marciana Marina, Isola d'Elba,
 Italy, 22-26 Jun 1998,
 arXiv:nucl-th/9901066.
\bibitem{Sarantsev:2005tg}
  A.V.~Sarantsev, V.A.~Nikonov, A.V.~Anisovich, E.~Klempt and U.~Thoma,
  Eur.\ Phys.\ J.\  A {\bf 25} (2005) 441.
\bibitem{Klempt:2002tt}
  E.~Klempt,
  Phys.\ Lett.\  B {\bf 559} (2003) 144.
\bibitem{Klempt:2007cp}
  E.~Klempt and A.~Zaitsev,
  Phys.\ Rept.\  {\bf 454} (2007) 1.
\bibitem{Lepage:1979za}
  G.~P.~Lepage and S.~J.~Brodsky,
  Phys.\ Rev.\ Lett.\  {\bf 43} (1979) 545
  [Erratum-ibid.\  {\bf 43} (1979) 1625].
\bibitem{Brodsky:2008pg}
  S.~J.~Brodsky and G.~F.~de Teramond,
  ``AdS/CFT and Light-Front QCD,''
 International School of Subnuclear Physics:
 45th Course: Searching for the ``Totally Unexpected" in
 the LHC Era, Erice, Sicily, Italy, 29 Aug - 7 Sep 2007,
  arXiv:0802.0514 [hep-ph].
\bibitem{Brodsky:2007hb}
  S.~J.~Brodsky and G.~F.~de Teramond,
  Phys.\ Rev.\  D {\bf 77} (2008) 056007.
\bibitem{Polchinski:2001tt}
  J.~Polchinski and M.~J.~Strassler,
  Phys.\ Rev.\ Lett.\  {\bf 88} (2002) 031601.
\bibitem{Klempt:2002vp}
  E.~Klempt,
  Phys.\ Rev.\  C {\bf 66} (2002) 058201.
\bibitem{Nambu:1978bd}
  Y.~Nambu,
  Phys.\ Lett.\  B {\bf 80} (1979) 372.
\bibitem{Baker:2002km}
  M.~Baker and R.~Steinke,
  Phys.\ Rev.\  D {\bf 65} (2002) 094042.
\bibitem{Baker:2003fn}
  M.~Baker,
  ``From QCD to dual superconductivity to effective string theory,''
 5th International Conference on Quark Confinement and the Hadron
 Spectrum, Gargnano, Brescia, Italy, 10-14 Sep 2002.
Published in *Gargnano 2002, Quark confinement and the hadron
spectrum* 204-210,
 arXiv:hep-ph/0301032.

\end{thebibliography}
\end{document}